\documentclass[11pt,a4paper]{article}
\pdfoutput=1
\usepackage{jheppub}

\usepackage{amsmath}
\usepackage{amssymb}
\usepackage{graphicx,color}

\usepackage{amsmath,amsthm}
\usepackage{txfonts}

\usepackage[sc]{mathpazo}
\usepackage[T1]{fontenc}

\usepackage[T1]{fontenc}

\usepackage{bm}

\newcommand{\Ref}[1]{(\ref{#1})}

\newtheorem{Theorem}{Theorem}[section]

\newcommand{\Z}{\mathbb{Z}}
\newcommand{\R}{\mathbb{R}}
\newcommand{\C}{\mathbb{C}}

\newcommand{\half}{\frac{1}{2}}

\newcommand{\Slc}{\mathrm{SL}(2,\mathbb{C})}

\newcommand{\Su}{\mathrm{SU}(2)}

\def\be{\begin{eqnarray}}
\def\ee{\end{eqnarray}}

\newcommand{\cf}{\mathcal F}

\newcommand{\ck}{\mathcal K}

\newcommand{\calr}{\mathcal R}
\newcommand{\cs}{\mathcal S}

\newcommand{\cv}{\mathcal V}
\newcommand{\cw}{\mathcal W}

\newcommand{\cz}{\mathcal Z}

\newcommand{\fs}{\mathfrak{s}}  \newcommand{\Fs}{\mathfrak{S}}

  \newcommand{\Fx}{\mathfrak{X}}

\renewcommand{\a}{\alpha}
\renewcommand{\b}{\beta}
\newcommand{\g}{\gamma}

\newcommand{\eps}{\varepsilon}

\newcommand{\sig}{\sigma}

\renewcommand{\l}{\lambda}

\renewcommand{\o}{\omega}

\renewcommand{\t}{\tau}

\newcommand{\rmd}{\mathrm d}

\newcommand{\lt}{\left}
\newcommand{\rt}{\right}

\newcommand{\lag}{\left\langle}
\newcommand{\rag}{\right\rangle}

\newcommand{\background}{(\mathring{j}_f,\mathring{g}_{ve},\mathring{z}_{vf})}
\newcommand{\sgn}{\mathrm{sgn}}
\newcommand{\vth}{\vartheta}

\title{Semiclassical Analysis of Spinfoam Model with a Small Barbero-Immirzi Parameter}

\author[]{Muxin Han}

\affiliation[]{Centre de Physique Th\'eorique%
\footnote{Unit\'e mixte de recherche (UMR 6207) du CNRS et des Universit\'es
de Provence (Aix-Marseille I), de la Meditarran\'ee (Aix-Marseille II) et du Sud (Toulon-Var); laboratoire affili\'e \`a la FRUMAM (FR 2291).}, CNRS-Luminy, Case 907, 13288 Marseille, France}

\emailAdd{Muxin.Han(AT)cpt.univ-mrs.fr} 

\abstract{We study the semiclassical behavior of Lorentzian Engle-Pereira-Rovelli-Livine (EPRL) spinfoam model, by taking into account of the sum over spins in the large spin regime. The large spin parameter $\l$ and small Barbero-Immirzi parameter $\g$ are treated as two independent parameters for the asymptotic expansion of spinfoam state-sum (such an idea was firstly pointed out in \cite{claudio}). Interestingly, there are two different spin regimes: $1\ll\g^{-1}\ll\l\ll\g^{-2}$ and $\l\geq\g^{-2}$. The model in two spin regimes has dramatically different number of effective degrees of freedom. In $1\ll\g^{-1}\ll\l\ll\g^{-2}$, the model produces in the leading order a functional integration of Regge action, which gives the discrete Einstein equation for the leading contribution. There is no restriction of Lorentzian deficit angle in this regime. In the other regime $\l\geq\g^{-2}$, only small deficit angle is allowed $|\Theta_f|\ll\g^{-1}\l^{1/2}$ mod $4\pi\Z$. When spins go even larger, only zero deficit angle mod $4\pi\Z$ is allowed asymptotically. In the transition of the two regimes, only the configurations with small deficit angle can contribute, which means one need a large triangulation in order to have oscillatory behavior of the spinfoam amplitude. }


\arxivnumber{}

\begin{document}

\maketitle


\section{Spinfoam Model and Scaling Parameters}

Loop Quantum Gravity (LQG) is an attempt to make a background independent, non-perturbative quantization of 4-dimensional General Relativity (GR) -- for reviews, see \cite{book,rev}. The discussion of the present paper concerns the covariant formulation of LQG, which is currently understood as the spinfoam formulation \cite{sfrevs}.  

Here we mainly focus on the semiclassical behavior of Lorentzian Engle-Pereira-Rovelli-Livine (EPRL) spinfoam model \cite{EPRL,Carlo} defined on an arbitrary simplicial complex. The semiclassical analysis is carried out by taking into account the sum over spins in the regime where all the spins are uniformly large. Such an analysis is a natural continuation of the previous studies of large spin asymptotics \cite{semiclassical,CF,HZ,hanPI}, which don't take into account the sum over spins. 

On the other hand, the result of the analysis also connects with the recent argument about the ``flatness problem'' proposed in \cite{flatness} when summing over spins is taken into account. In \cite{flatness} the authors argue that the sum over spins in the spinfoam model may impose a projection at least in the semiclassical level, which projects out a large amount of nontrivial (semi-)classical simplicial geometry, and leaves only the geometry with deficit angle $\Theta_f=0$ mod $4\pi\Z$. 

The analysis in the present paper treats more carefully the semiclassical analysis of the spin-sum in the large spin regime. The method of perturbative expansion is employed in the large spin regime, where the spin-scaling $\l$ is a natural expansion parameter. However an additional scaling parameter has to be introduced in order to evade the flatness problem mentioned above. The resulting expansion uses the combinations of the two scaling parameters. A first idea of the additional parameter leads to the Barbero-Immirzi parameter $\g$. The idea of considering $\g$ to be an additional scaling parameter in the semicalssical analysis is firstly proposed in \cite{claudio}. It is shown in the following that the perturbative expansion from such an idea works in the regime $1\ll\g^{-1}\ll\l\ll\g^{-2}$. It gives as the leading contribution a functional integration of Regge action, where a discrete Einstein equation is reproduced. However as the spin-scaling $\l\geq g^{-2}$, we have to define a scaling of deficit angle, which is treated as the additional expansion parameter. We show explicitly how the additional scaling parameter can be developed with or without Barbero-Immirzi parameter. The resulting expansion in the regime $\l\geq g^{-2}$ may be viewed as a curvature expansion, where only small deficit angle is allowed $\Theta_f\sim o(\g^{-1}\l^{1/2})$ mod $4\pi k_f$ ($k_f\in\Z$). At least in the $k_f=0$ branch, the leading contribution to the spinfoam amplitude gives a functional integration of Regge action with only small deficit angle contributions. When spin-scaling $\l\to\infty$, only zero deficit angle mod $4\pi\Z$ is allowed asymptotically, which reproduces the flatness result argued in \cite{flatness}.

Interestingly, the EPRL spinfoam model in two spin regimes $1\ll\g^{-1}\ll\l\ll\g^{-2}$ and $\l\geq g^{-2}$ has dramatically different number of effective degrees of freedom. In $1\ll\g^{-1}\ll\l\ll\g^{-2}$, the effective degrees of freedom contain the Lorentzian geometries with arbitrary values of deficit angle. But in $\l\geq g^{-2}$ the effective degrees of freedom only admit the Lorentzian geometries with the deficit angle bounded by $|\Theta_f|\ll\g^{-1}\l^{1/2}$ mod $4\pi\Z$. The situation is illustrated in FIG.\ref{phases}.

As the starting point of the analysis in this paper, we employ the following path integral representation, proposed in \cite{hanPI}, for the EPRL spinfoam state-sum model on a simplical complex $\ck$:
\be
A(\ck)&=&\sum_{J_{f}}d_{J_f} \int_{\Slc}\prod_{\left( v,e\right)}
\rmd g_{ve} \int_{\mathbb{CP}^{1}}\prod_{v\in \partial f} \rmd{z_{vf}}\ e^{S[J_f,g_{ve},z_{vf}]}
\ee
where $f$ labels a triangle in the simplicial complex $\ck$ or a dual face in the dual complex $\ck^*$, $e$ labels a tetrahedron in $\ck$ or an edge in $\ck^*$, and $v$ labels a 4-simplex in $\ck$ or a dual vertex in $\ck^*$.  $J_f$ labels the SU(2) irreps associated to each triangle. $d_J$ is the dimension of the SU(2) irrep with spin $J$. $g_{ve}$ is a $\Slc$ group variable associated with each dual half-edge.  $z_{vf}$ is a 2-component spinor. The integrand written into an exponential form $e^S$ with the spinfoam action $S$ written as
\be
S[J_f,g_{ve},z_{vf}]
&=&\sum_{(e,f)}\lt[J_f\ln\frac{\lag Z_{vef},Z_{v'ef}\rag^{2}}{ \lag Z_{v'ef},Z_{v'ef}\rag \lag Z_{vef},Z_{vef}\rag}+i\g J_f \ln \frac{\lag Z_{vef},Z_{vef}\rag}{\lag Z_{v'ef},Z_{v'ef}\rag}\rt]
\label{action}
\ee
where $Z_{vef}=g_{ve}^\dagger z_{vf}$ and $\g\in \R$ is the Barbero-Immirzi paramter. We refer to \cite{hanPI} for a derivation of such a path integral representation. The spinfoam action $S$ has the following discrete gauge symmetry: Flipping the sign of individual group variable $g_{ve}\mapsto -g_{ve}$ leaves $S$ invariant. Thus the space of group variable is essentially the restricted Lorentz group $\mathrm{SO}^+(1,3)$ rather than its double-cover $\Slc$. $S$ also has the continuous gauge degree of freedom: 
(1) Rescaling of each $z_{vf}$\footnote{The measure $\rmd z_{vf}$ is a scaling invariant measure on $\mathbb{CP}^1$.}, $z_{vf}\mapsto \l z_{vf}, \l\in\C\setminus\{0\}$; 
(2) $\Slc$ gauge transformation at each vertex $v$, $g_{ve}\mapsto x^{-1}_vg_{ve}, z_{vf}\mapsto x^\dagger_vz_{vf}, x_v\in\Slc$; 
(3) SU(2) gauge transformation on each edge $e$, $g_{ve}\mapsto g_{ve}h_e^{-1}, h_e\in\Su$.

For the convenience of the discussion, we define the notion of the partial-amplitude $A_{j_f}(\ck)$ by collecting all the integrations
\be
A_{J_f}(\ck):=\int
\rmd g_{ve} \int\rmd z_{vf}\ e^{S\lt[J_f,g_{ve},z_{vf}\rt]}  \label{Aj}
\ee
So that the spinfoam state-sum model is given by a sum of partial amplitude over all the spin configurations $\{J_f\}_f$ on the simplicial complex $\ck$ 
\be
A(\ck)=\sum_{J_f}d_{J_{f}}A_{J_f}(\ck).\label{sumJ}
\ee
Note that the infinite spin-sum in $A(\ck)$ may result in a divergent result. A way to regularizing the spin-sum is to replace $\Slc$ in the definition by the quantum group $\mathrm{SL}_q(2,\C)$ \cite{QSF}, which also relates to the cosmological constant term in spinfoam formulation \cite{QSFasymptotics}. 

Apart from the quantum group regulator $q$, the Barbero-Immirzi parameter $\g$ is the only free parameter entering the definition of spinfoam model $A(\ck)$. In the present analysis we assume a small Barbero-Immirzi parameter $\g\ll1$.

So far the semiclassical properties of the spinfoam model is mostly understood in the large spin regime of the spinfoam state-sum Eq.\Ref{sumJ} (see e.g. \cite{hanPI,statesum,HZ,CF,semiclassical}). In the large spin regime, all the spins $J_f$ are uniformly large, so we can define a scaling $J_f=\l j_f$, where $\l\ll1$ is a large parameter to scale the spins uniformly and $j_f\sim o(1)$. In terms of the scaling paramter $\l$, the spinfoam action in the large spin regime is written as the following:
\be
\l S[j_f,g_{ve},z_{vf}]=\sum_{(e,f)}\lt[\l j_f\ln\frac{\lag Z_{vef},Z_{v'ef}\rag^{2}}{ \lag Z_{v'ef},Z_{v'ef}\rag \lag Z_{vef},Z_{vef}\rag}+i\g\l j_f \ln \frac{\lag Z_{vef},Z_{vef}\rag}{\lag Z_{v'ef},Z_{v'ef}\rag}\rt].\label{Sj}
\ee
Since $\g$ is a free paramter, the above expression of spinfoam action suggests that we can define another scaling parameter $\b$ by
\be
\b=\l\g,
\ee
So that we can write
\be
&&\l S[j_f,g_{ve},z_{vf}]=\sum_{f}\Big(\l j_f\cv_f\lt[g_{ve},z_{vf}\rt] +i\b j_f \ck_f\lt[g_{ve},z_{vf}\rt]\Big)\nonumber\\
\text{where}\ \ \ &&\cv_f\lt[g_{ve},z_{vf}\rt]\equiv \sum_e\ln\frac{\lag Z_{vef},Z_{v'ef}\rag^{2}}{ \lag Z_{v'ef},Z_{v'ef}\rag \lag Z_{vef},Z_{vef}\rag}\ \ \text{and}\ \ \ck_f\lt[g_{ve},z_{vf}\rt]=\sum_e\ln \frac{\lag Z_{vef},Z_{vef}\rag}{\lag Z_{v'ef},Z_{v'ef}\rag}
\ee
The scaling parameter $\b$ can be considered as an independent parameter for the expansion since it can be considered as an reparametrization of the paramter space defined by $(\l,\g)$. When we consider the path integral representation of the partial amplitude 
\be
A_{j_f,\l,\b}(\ck) =\int\rmd g_{ve} \int\rmd z_{vf}\ e^{\l \sum_f j_f\cv_f\lt[g_{ve},z_{vf}\rt]}e^{i\b \sum_{f}j_f \ck_f\lt[g_{ve},z_{vf}\rt]},\label{Ajlb}
\ee
The stationary phase analysis can be applied to the first exponential in the integrand, in order to obtain an $\l^{-1}$-expansion, while the second exponential are simply evaluated at the critical points given by the first exponential. Such an idea has been proposed in the early work in \cite{claudio}, and is motivated by the spinfoam graviton propagator computation \cite{propagator,3pt}. The advantage of such a procedure is the following: It turns out (in the analysis of the next section) that the ``potential'' $\cv_f\lt[g_{ve},z_{vf}\rt]$ gives exactly the same set of critical points as the one given by the full spinfoam action $S$, classified in \cite{hanPI,HZ}\footnote{We thank the private communication with E. Bianchi at this point.}. At the critical spinfoam configurations corresponding to nondegenerate simplical geometries, the potential $\cv_f\lt[g_{ve},z_{vf}\rt]$ vanishes at the ``time-oriented'' configurations (defined in \cite{hanPI}), and equals $\pi$ otherwise, while the critical value of $\sum_{f}j_f \ck_f\lt[g_{ve},z_{vf}\rt]$ gives the Regge action evaluated at the corresponding nondegenerate geometry, where the critical value of $\ck_f\lt[g_{ve},z_{vf}\rt]$ is the deficit angle at $f$. It turns out that $\g\ll1$ and making $\b$ and $\l$ as two independent expansion parameter results in the nondecaying (perturbative) state-sum amplitude with a finite deficit angle, while if $\g\sim o(1)$, the expansion in $\l^{-1}$ in \cite{statesum} shows that the state-sum amplitude decays exponentially unless the deficit angle is smaller than $o(\l^{-1/2})$. However as it is shown in the following, the expansion procedure decribed above by making $\l,\b$ independent is only consistent in a certain regime of the spin-sum $\sum_J$. Beyond such a regime this procedure has to be replaced by the expansion in \cite{statesum}.

\section{Large-$\l$ Asymptotic Expansion}

\subsection{Critical Configurations}\label{criticalconfigurations}

In order to analyze the (perturbative) semiclassical behavior of the spinfoam state-sum, a preliminary step is the asymptotic analysis of the partial amplitude $A_{j_f,\l,\b}(\ck)$ as an asymptotic expansion in $\l^{-1}$. It is guided by the following general result (Theorem 7.7.5 in \cite{stationaryphase}):

\begin{Theorem}\label{asymptotics}
Let $K$ be a compact subset in $\R^N$, $X$ an open neighborhood of $K$, and $k$ a positive integer. If (1) the complex functions $u\in C^{2k}_0(K)$, $S\in C^{3k+1}(X)$ and $\Re(S)\leq 0$ in $X$; (2) there is a unique point $x_0\in K$ satisfying $\Re(S)(x_0)=0$, $S'(x_0)=0$, and $\det S''(x_0)\neq 0$. $S'\neq0$ in $K\setminus \{x_0\}$, then we have the following estimation:
\be
\int_K u(x)e^{\l S(x)}\rmd x=e^{\l S(x_0)}\lt(\frac{2\pi}{\l}\rt)^{\frac{N}{2}}\frac{e^{\mathrm{Ind}(S'')(x_0)}}{\sqrt{\det(S'')(x_0)}}\sum_{s=0}^{\infty}\lt(\frac{1}{\l}\rt)^s L_s u(x_0)
\ee
$L_s u(x_0)$ is a differential operator of order $2s$ acting on $u(x)$:
\be
L_s u(x_0)=i^{-s}\sum_{l-m=s}\sum_{2l\geq 3m}\frac{2^{-l}}{l!m!}\lt[\sum_{a,b=1}^NH^{-1}_{ab}(x_0)\frac{\partial^2}{\partial x_a\partial x_b}\rt]^l\lt(g_{x_0}^m u\rt)(x_0)\label{Lsu}
\ee
where $H(x)=S''(x)$ denotes the Hessian matrix and the function $g_{x_0}(x)$ is given by
\be
g_{x_0}(x)=S(x)-S(x_0)-\frac{1}{2}H^{ab}(x_0)(x-x_0)_a(x-x_0)_b
\ee
such that $g_{x_0}(x_0)=g_{x_0}'(x_0)=g_{x_0}''(x_0)=0$.

\end{Theorem}

For each $s$, $L_s$ is a differential operator of order $2s$ acting on $u(x)$. For example we list the possible types of terms in the sums corresponding to $s=1$ and $s=2$
\begin{itemize}

\item In the case $s=1$, the possible $(m,l)$ are $(m,l)=(0,1),(1,2),(2,3)$ to satisfy $2l\geq 3m$. The corresponding terms are of the types
\be
(m,l)=(0,1):&& \partial^2u(x_0)\nonumber\\
(m,l)=(1,2):&& \partial^3g_{x_0}(x_0)\partial u(x_0),\ \ \partial^4g_{x_0}(x_0)u(x_0) \nonumber\\
(m,l)=(2,3):&&\partial^3g_{x_0}(x_0)\partial^3g_{x_0}(x_0) u(x_0)
\ee
where the indices of $\partial$ are contracted with the Hessian matrix $H(x_0)$.

\item In the case $s=2$, the possible $(m,l)$ are $(m,l)=(0,2),(1,3),(2,4),(3,5),(4,6)$ to satisfy $2l\geq 3m$. The corresponding terms are of the types
\be
(m,l)=(0,2):&& \partial^4u(x_0)\nonumber\\
(m,l)=(1,3):&& \partial^pg_{x_0}(x_0)\partial^q u(x_0),\ \ \ \
(p\geq 3,\ p+q=6) \nonumber\\
(m,l)=(2,4):&& \partial^{p_1}g_{x_0}(x_0)\partial^{p_2}g_{x_0}(x_0) \partial^qu(x_0)\ \ \ \
(p_1,p_2\geq 3.\ p_1+p_2+q=8)\nonumber\\
(m,l)=(3,5):&& \partial^{p_1}g_{x_0}(x_0)\partial^{p_2}g_{x_0}(x_0)\partial^{p_3}g_{x_0}(x_0) \partial^qu(x_0)\ \ \ \  (p_1,p_2,p_3\geq 3.\ p_1+p_2+p_3+q=10)\nonumber\\
(m,l)=(4,6):&& \partial^{3}g_{x_0}(x_0)\partial^{3}g_{x_0}(x_0)\partial^{3}g_{x_0}(x_0)\partial^{3}g_{x_0}(x_0) u(x_0)
\ee
where the indices of $\partial$ are contracted with the Hessian matrix $H(x_0)$.

\end{itemize}

We apply the above stationary phase approximation to the integral $A_{j_f,\l,\b}(\ck)$ in Eq.\Ref{Ajlb} in order to obtain an $\l^{-1}$ expansion. Recall that the second exponential in the integrand of Eq.\Ref{Ajlb} doesn't depend on $\l$, thus the asymptotic expansion is determined by the critical points given by the potential 
\be
\l \sum_fj_f\cv_f\lt[g_{ve},z_{vf}\rt]=\l \sum_{(e,f)}j_f\ln\frac{\lag Z_{vef},Z_{v'ef}\rag^{2}}{ \lag Z_{v'ef},Z_{v'ef}\rag \lag Z_{vef},Z_{vef}\rag}
\ee

The critical points of the above potential are give by the solutions of the critical equations $\Re(\cv_f)=\delta_{g_{ve}}\cv_f=\delta_{z_{vf}}\cv_f=0$. The derivation of these critical equations follows in the same way as the ones derived for the spinfoam action $S$ in \cite{hanPI}, by simply setting $\g=0$. We skip the derivation here and list the resulting critical equations, which are exactly the same as the critical equations from $S$:
\be
\Re(\cv_f)=0:&& \frac{g^\dagger_{ve}z_{vf}}{\lt|\lt|Z_{vef}\rt|\rt|}=e^{i\a^f_{vv'}}\frac{g^\dagger_{v'e}z_{v'f}}{\lt|\lt|Z_{v'ef}\rt|\rt|},\label{1}\\
\delta_{z_{vf}}\cv_f=0:&& \frac{g_{ve}g_{ve}^\dagger z_{vf}}{\lag Z_{vef},Z_{vef}\rag}=\frac{g_{ve'}g_{ve'}^\dagger z_{vf}}{\lag Z_{ve'f},Z_{ve'f}\rag},\label{2}\\
\delta_{g_{ve}}\cv_f=0:&& \sum_{f}j_{f}\eps_{ef}(v)\frac{ \lag Z_{vef}\ \vec{\sig}\ Z_{vef}\rag }
{\lag Z_{vef},Z_{vef}\rag}=0.\label{3}
\ee
where $\a^f_{vv'}$ is an arbitrary phase, and the incidence matrix $\eps_{ef}(v)$ is given by
\begin{equation}
\eps_{ef}(v)=
\begin{cases}
\:\:\:0&\text{if $v\notin\partial f$}\cr
\:\:\:1&\text{if $v=t(e)$ with $e\in\partial f$}\cr
-1&\text{if $v=s(e)$ with $e\in\partial f$}
\end{cases}
\end{equation}
$\eps_{ef}(v)$ satisfies the following relations:
\be
\eps_{ef}(v)=-\eps_{e'f}(v)\ \ \ \ \text{and}\ \ \ \ \eps_{ef}(v)=-\eps_{ef}(v').
\ee

Because the critical equations from the potential $\cv_f$ is identical to the critical equations from the spinfoam action $S$, the geometrical interpretations of the critical configurations follows in the same way as it was developed in \cite{hanPI,HZ}. The results is summarized in the following (see \cite{hanPI,HZ} for details):
\begin{itemize}

\item The most interesting class of critical configurations satisfies an additional nondegeneracy condition at each vertex $v$:
\be
\prod_{e_1,e_2,e_3,e_4=1}^5\det\Big(N_{e_1}(v),N_{e_2}(v),N_{e_3}(v),N_{e_4}(v)\Big)\neq0\label{proddet}
\ee
where $N_e(v):=g_{ve}(1,0,0,0)^t$. A critical configuration $(j_f,g_{ve},z_{vf})$ satisfying Eq.\Ref{proddet} is 1-to-1 corresponding to a geometrical data $(\pm_vE_\ell(v),\eps)$ where $E_\ell(v)$ is an edge-vector associated to each edge $\ell$ of the simplicial complex $\ck$. The set of $E_\ell(v)$ is called a discrete cotetrad, and determines a nondegenerate simplicial Lorentzian geometry on $\ck$, with nonzero oriented volume $V_4(v)$ of each geometrical 4-simplex. The geometrical area of each triangle is given by $\g j_f$. $\eps=\pm1$ is a global sign on the simplicial complex $\ck$, which is determined by the boundary data if $\ck$ has a boundary. $\pm_v$ labels a sign ambiguity at each $v$ in relating the cotetrad to the spinfoam critical data.

\item In case that the critical configuration $(j_f,g_{ve},z_{vf})$ violates the nondegeneracy condition Eq.\Ref{proddet}, it doesn't admit a geometrical interpretation as nondegenerate Lorentz geometry. However a subclass of such critical configurations admits the interpretation as nondegenerate Euclidean geometries. More precisely there is an 1-to-1 correspondence between a critical configuration of such a type and a set of geometrical data $(\pm_vE^E_\ell(v),\eps,\eps_e(v))$, where $E^E_\ell(v)$ is a discrete cotetrad for Euclidean geometry, and $\eps_e(v)$ is a sign associated to each pair $(e,v)$.

\item The rest of the critical configurations violating Eq.\Ref{proddet} only correspond to degenerate geometries, which are called vector geometries. The geometrical data of a vector geometry is a set of 3-vectors associated to the triangles.

\end{itemize} 

Now let's consider the functions $\cv_f[g_{ve},z_{vf}]$ and $\ck_f[g_{ve},z_{vf}]$ evaluated at the critical configurations: When the critical configuration satisfies the nondegeneracy condition Eq.\Ref{proddet}, the spinfoam loop holonomy $G_f(v)$ along the boundary of the dual face $f$ can be computed at the critical configuration \cite{HZ} ($g_{ev}=g_{ve}^{-1}$):
\be
G_f(v)=\overleftarrow{\prod_{e\in\partial f}}g_{v'e}g_{ev}=\exp\lt[\frac{* E_{\ell_1}(v)\wedge E_{\ell_2}(v)}{|*E_{\ell_1}(v)\wedge E_{\ell_2}(v)|}\sgn(V_4)\Theta_f+\frac{E_{\ell_1}(v)\wedge E_{\ell_2}(v)}{|E_{\ell_1}(v)\wedge E_{\ell_2}(v)|}\pi n_f\rt]
\ee
where the continuous parameter $\Theta_f$ is interpreted as the deficit angle hinged by the trangle $f$, and the discrete parameter $n_f$ can be either 0 or 1. Note that we have assumed $\sgn(V_4)$ is a constant along the loop. It is shown in \cite{HZ} that the critical values of $\cv_f[g_{ve},z_{vf}]$ and $\ck_f[g_{ve},z_{vf}]$ relates respectively to the parameter $\Theta_f$ and $n_f$:
\be
\cv_f[g_{ve},z_{vf}]=i\eps\pi n_f\equiv i\a_f\ \ \ \text{and}\ \ \ \ck_f[g_{ve},z_{vf}]=\eps\ \sgn(V_4)\Theta_f. 
\ee 

In case that the critical configuration $(j_f,g_{ve},z_{vf})$ violates the nondegeneracy condition Eq.\Ref{proddet}, the analysis in \cite{HZ} shows that the critical value of $\ck_f[g_{ve},z_{vf}]$ vanishes identically and the critical value of $\cv_f[g_{ve},z_{vf}]$ is given by
\be
\cv_f[g_{ve},z_{vf}]=i\eps \lt[\sgn(V_4)\Theta^E_f+\pi n_f\rt]\ \ \ \text{or}\ \ \ \cv_f[g_{ve},z_{vf}]=i\Phi_f
\ee
for the Euclidean geometrical interpretation or the vector geometrical interpretation, where the continuous parameter $\Theta^E_f$ is the Euclidean deficit angle, the discrete parameter $n_f=0,1$, and $\Phi_f$ is the vector geometry angle, as a continuous parameter. 

Given a critical configuration $(j_f,g_{ve},z_{vf})$ corresponding to a nondegenerate Lorentzian geometry, it is called globally oriented if $\sgn(V_4)$ is a constant everywhere on the simplicial complex, and it is called time-oriented if $n_f=0$ for all $f$. The condition for the time-oriented critical configuration requires the loop holonomy of the spin connection compatible with the cotetrad $E_\ell(v)$ should belong to the restricted Lorentz group $\mathrm{SO}^+(1,3)$ \cite{hanPI}.  

The following table summarizes the critical values of $\cv_f[g_{ve},z_{vf}]$ and $\ck_f[g_{ve},z_{vf}]$ at different type of critical configurations. In the following we often denote the critical values of $\cv_f$ by $i\alpha_f$ and the critical value of $\ck_f$ by $\vth_f$.
\begin{center}
\begin{tabular}{| l || c | r |}
\hline
  & $\cv_f\equiv i\a_f$ & $\ck_f\equiv\vth_f$\ \ \ \\ \hline \hline
\ Lorentzian Time-Oriented&\ \ 0& $\eps\ \sgn(V_4)\Theta_f$\  \\ \hline
Lorentzian Time-Unoriented&\ \ $i\eps\pi$ & $\eps\ \sgn(V_4)\Theta_f$\  \\ \hline
\ \ \ \ \ \ \ \ \ \ \ Euclidean & $i\eps \lt[\sgn(V_4)\Theta^E_f+\pi n_f\rt]$ & 0\ \ \ \ \ \ \ \ \\ \hline
\ \ \ \ \ \ \ \ \ \ \ \ \ Vector & $i\Phi_f$ &  0\ \ \ \ \ \ \ \  \\ \hline
  \end{tabular}\\
  \vspace{0.4cm}
  \underline{Table 1}.
\end{center}

\subsection{$\l^{-1}$-Expansion} \label{lexp}

Recall the integration representation of partial amplitude
\be
A_{j_f,\l,\b}(\ck) =\int\rmd g_{ve} \int\rmd z_{vf}\ e^{\l \sum_f j_f\cv_f\lt[g_{ve},z_{vf}\rt]}e^{i\b \sum_{f}j_f \ck_f\lt[g_{ve},z_{vf}\rt]}
\ee
In order to derive an asymptotic expansion for the above integral for all values of $j_f$, we should treat $j_f$ as a parameter and employ the generalized stationary phase analysis with parameter, which is also known as almost-anaytic machinery \cite{almostanalytic}:

\begin{Theorem} \label{almostanalytic}

Let $S(j,x)$, $j\in\R^k,\ x\in\R^N$, be an smooth function in a neighborhood of $(\mathring{j},\mathring{x})$. We suppose that $\Re\lt[S(j,x)\rt]\leq0$, $\Re \lt[S(\mathring{j},\mathring{x})\rt]=0$, $\delta_x S(\mathring{j},\mathring{x})=0$, and $\delta^2_{x,x}S(\mathring{j},\mathring{x})$ is nondegenerate. We denote by $\cs(j,z)$, $j\in\C^k,\ z=x+iy\in\C^n$ an (nonunique) almost-analytic extension\footnote{An almost analytic extension $\tilde{f}$ of $f\in C^\infty(\R)$ in a neighborhood $\o$ satisfies (1) $\tilde{f}=f$ in $\o\cap\R$, (2) $|\partial_{\bar{z}}\tilde{f}|\leq C_N|\Im(z)|^N$ for all $N\in \Z_+$, i.e. $\partial_{\bar{z}}\tilde{f}$ vanishes to infinite order on the real axis. } of $S(j,x)$ to a complex neighborhood of $(\mathring{j},\mathring{x})$. The equations of motion $\delta_z \cs=0$ define an almost-analytic manifold $M$ in a neighborhood of $(\mathring{j},\mathring{x})$, which is of the form $z=Z(j)$. On $M$ and inside the neighborhood, there is a positive constant $C$ such that for $j\in\R^k$
\be
-\Re[\cs(j,z)]\geq C|\Im(z)|^2,\ \ \ \ z=Z(j) \label{recs}
\ee
We have the following asymptotic expansion for the integral
\be
\int e^{\l S(j,x)}\ u(x)\ \rmd x\sim e^{\l \cs\lt[j,Z(j)\rt]}\lt(\frac{1}{\l}\rt)^{\frac{N}{2}}\sqrt{\det\lt(\frac{2\pi i}{ \cs''\lt[j,Z(j)\rt]}\rt)}\sum_{s=0}^\infty\lt(\frac{1}{\l}\rt)^{s}\lt[L_{s}\tilde{u}\rt]\Big(Z(j)\Big)\label{expcs}
\ee
where $u(x)\in C^\infty_0(K)$ is a compact support function on $K$ inside the domain of integration. $N$ is the number of independent of $x$-variables, the same as the number of holomorphic $z$-variables. The differential operator $L_{s}$ is defined in the same way as in Theorem \ref{asymptotics} but operates on an almost analytic extension $\tilde{u}(z)$ of $u(x)$ and evaluating the result at $z=Z(j)$. The branch of the square-root is defined by requiring $\sqrt{\det\lt({2\pi i}\big/{ \cs''\lt[j,Z(j)\rt]}\rt)}$ to deform continuously to $1$ under the homotopy:
\be
(1-s)\frac{2\pi i}{ \cs''\lt[j,Z(j)\rt]}+s\mathbf{I}\in\mathrm{GL}(n,\C),\ \ \ \ s\in[0,1].
\ee
Note that the asymptotic expansions from two different almost-analytic extensions of the pair $S(j,x),u(x)$ are different only by an contribution bounded by $C_K\l^{-K}$ for all $K\in\Z_+$.

\end{Theorem}

The (almost)-analytic extension of the spinfoam action is given in \cite{statesum}, where the extended spinfoam action $\cs$ depends on a pair of group variables $(g_{ve},\tilde{g}_{ve})\in\Slc\times\Slc$ and a pair of spinors $(z_{vf},\tilde{z}_{vf})\in\mathbb{CP}^1\times\mathbb{CP}^1$. The expression of $\cs$ is given explicitly by
\be
\l\cs\lt[j_f,g_{ve},\tilde{g}_{ve},z_{vf},\tilde{z}_{vf}\rt]&=&\sum_f\Big[\l j_f\tilde{\cv}_f\lt[g_{ve},\tilde{g}_{ve},z_{vf},\tilde{z}_{vf}\rt]+i\b j_f\tilde{\ck}_f\lt[g_{ve},\tilde{g}_{ve},z_{vf},\tilde{z}_{vf}\rt]\Big]\nonumber\\
\text{where}\ \ \ \ 
\tilde{\cv}_f\lt[g_{ve},\tilde{g}_{ve},z_{vf},\tilde{z}_{vf}\rt]&=&\sum_{e\subset f}\ln\frac{\lt(g_{ve}^tz_{vf}\cdot\tilde{g}_{v'e}^t\tilde{z}_{v'f}\rt)^{2}}{ \lt( g_{v'e}^tz_{v'f}\cdot\tilde{g}_{v'e}^t\tilde{z}_{v'f}\rt) \lt(g_{ve}^tz_{vf}\cdot\tilde{g}_{ve}^t\tilde{z}_{vf}\rt)}\nonumber\\
\text{and}\ \ \ \ 
\tilde{\ck}_f\lt[g_{ve},\tilde{g}_{ve},z_{vf},\tilde{z}_{vf}\rt]&=&\sum_{e\subset f}\ln \frac{\lt( g_{ve}^tz_{vf}\cdot\tilde{g}_{ve}^t\tilde{z}_{vf}\rt)}{\lt( g_{v'e}^tz_{v'f}\cdot\tilde{g}_{v'e}^t\tilde{z}_{v'f}\rt)}\label{acaction}
\ee
The functions $\tilde{\cv}_f$ and $\tilde{\ck}_f$ are actually analytic functions in a neighborhood of a critical configuration $(j_f,g_{ve},z_{vf})$. The equations of motion are given by $\delta_{g}\tilde{\cv}_f=\delta_{\tilde{g}}\tilde{\cv}_f=\delta_{z}\tilde{\cv}_f=\delta_{\tilde{z}}\tilde{\cv}_f=0$, which defines an analytic manifold $Z(j)=\lt(g_{ve}(j),\tilde{g}_{ve}(j),z_{vf}(j),\tilde{z}_{vf}(j)\rt)$ modulo gauge transformations. The gauge transformations of $\cs$ are classified in \cite{statesum}.

By apply the above theorem formally to the spinfoam action $\cs$, we obtain the following $\l^{-1}$-expansion in a neighborhood $K$ at a critical configuration $(j_f,g_{ve},z_{vf})$:
\be
A_{j_f,\l,\b}\sim  e^{\l \sum_fj_f\tilde{\cv}_f\lt[Z(j)\rt]}\lt(\frac{1}{\l }\rt)^{\frac{N_{g,z}}{2}}\sqrt{\det\lt(\frac{2\pi i}{ \sum_fj_f\tilde{\cv}_f''\lt[Z(j)\rt]}\rt)}\sum_{s=0}^\infty\lt(\frac{1}{\l}\rt)^{s}\lt[L_{s}\tilde{u}\rt]\Big(j,Z(j)\Big)
\ee
where $N_{g,z}$ is the number of degree of freedom in the holomorphic variables $g_{ve},\tilde{g}_{ve},z_{vf},\tilde{z}_{vf}$ modulo gauge transformations. Here $\tilde{u}$ is given by
\be
\tilde{u}\lt[j_f,g_{ve},\tilde{g}_{ve},z_{vf},\tilde{z}_{vf}\rt]=e^{i\b\sum_fj_f\tilde{\ck}_f\lt[g_{ve},\tilde{g}_{ve},z_{vf},\tilde{z}_{vf}\rt]}\tilde{\mu}\lt[g_{ve},\tilde{g}_{ve},z_{vf},\tilde{z}_{vf}\rt]
\ee
where $\tilde{\mu}$ contains the almost-analytic extensions of the Jacobian of the integral measure (with respect to the Lebesgue measure) and a compact support test function supported on the neighborhood $K$\footnote{We can decompose in general the integral on real space $\int\rmd\mu(x)e^{\l S(x)}=\sum_I\int\rmd\mu(x)e^{\l S(x)}u_I(x)$, where each $u_I(x)$ is compact support on $K_I$ and $\sum_Iu_I(x)=1$ (a partition of unity). Each $K_I$ only contains a single critical point. See \cite{stationaryphase} for details. }.  

Eq.\Ref{recs} implies that
\be
\Re\lt(\sum_fj_f\tilde{\cv}_f\lt[Z(j)\rt]\rt)\leq -C\lt|\Im\lt(Z(j)\rt)\rt|^2\label{real}
\ee
is non-positive.

Recall Theorem \ref{asymptotics} that $L_s$ acting on $\tilde{u}$ is a differential operator (with respect to the variables $g_{ve},\tilde{g}_{ve},z_{vf},\tilde{z}_{vf}$) of order $2s$, thus $L_{s}\tilde{u}$ can be written as
\be
\lt[L_{s}\tilde{u}\rt]\Big(Z(j)\Big)=e^{i\b\sum_fj_f\tilde{\ck}_f\lt[Z(j)\rt]}\sum_{r=0}^{2s}\b^r f_{r,s}\Big(j,Z(j)\Big)
\ee
where $r$ is the number of derivatives acting on the exponential. Therefore we obtain the following expansion of the partial amplitude $A_{j_f,\l,\b}(\ck)$ in the neighbourhood at a critical configuration:
\be
A_{j_f,\l,\b}\sim  e^{\l \sum_fj_f\tilde{\cv}_f\lt[Z(j)\rt]}e^{i\b\sum_fj_f\tilde{\ck}_f\lt[Z(j)\rt]}\lt(\frac{1}{\l }\rt)^{\frac{N_{g,z}}{2}}\sqrt{\det\lt(\frac{2\pi i}{ \sum_fj_f\tilde{\cv}_f''\lt[Z(j)\rt]}\rt)}\sum_{s=0}^\infty\sum_{r=0}^{2s}\lt(\frac{\b^{r}}{\l^s}\rt)f_{r,s}\Big(j,Z(j)\Big)\label{VK}
\ee
It is clear from the expression that the above asymptotic expansion makes sense only when $\frac{\b^{r}}{\l^s}\ll1, \forall r\leq2s$, i.e.
\be
\b\ll\sqrt{\l}\ \ \ \Rightarrow\ \ \ \g\ll\frac{1}{\sqrt{\l}}\ \ \text{or}\ \ \l\ll\frac{1}{\g^2}
\ee
since $\b=\g\l$ by definition. The above analysis concerns the large spin regime ($\l\gg1$) of the spinfoam state-sum, so such an asymptotic expansion exists only when the Barbero-Immirzi parameter is small $\g\ll 1$. In case such an expansion exists, it only valid when the large spin is bounded by $\l\ll\frac{1}{\g^2}$.

We can define an effective action for the partial amplitude $A_{j_f,\l,\b}(\ck)$, which we call the spin effective action $W_\ck[j_f,\l,\b]$:
\be
A_{j_f,\l,\b}(\ck)=\exp W_\ck[j_f,\l,\b]\ \ \ \text{so}\ \ \ A(\ck)=\sum_{j_f}d_{J_f}\exp W_\ck[j_f,\l,\b]
\ee
As far as $1\ll\l\ll\g^{-2}$ is satisfied, by the above asymptotic analysis of the partial amplitude, the expression of the spin effective action can be written as
\be
W_\ck[j_f,\l,\b]&=&\l \sum_fj_f\tilde{\cv}_f\lt[Z(j)\rt]+i\b\sum_fj_f\tilde{\ck}_f\lt[Z(j)\rt]-\frac{N_{g,z}}{2}\ln\l+\half\ln\det\lt(\frac{2\pi i}{ \sum_fj_f\tilde{\cv}_f''\lt[Z(j)\rt]}\rt)\nonumber\\
&&+\ o\lt(\frac{\b^r}{\l^s}\rt)_{r\leq 2s}.\label{Wexpression}
\ee

In the following it turns out that $\b=\g\l$ has to be a large parameter, such that $\b^{-1}$ is another expansion parameter. Therefore $1\ll\g^{-1}\ll \l\ll\g^{-2}$ is required in the following analysis.

\section{Spin-Sum in the Regime $\g^{-1}\ll\l\ll\g^{-2}$}

\subsection{Implementation of Spin-Sum}\label{implementation}

In this section we take into account the spin-sum in $A(\ck)=\sum_{J_f}d_{J_f}A_{J_f}(\ck)$ in the regime $\g^{-1}\ll\l\ll\g^{-2}$. By the above asymptotic power-series expansion of the spin effective action $W_\ck[j_f,\l,\b]$, we can write the spin-sum by
\be
A(\ck)=\sum_{j_f=-\infty}^\infty d_{J_f}\t(j_f)\exp W_\ck[j_f,\l,\b]=\lt(\frac{1}{\l}\rt)^{\frac{N_{g,z}}{2}}\sum_{J_f\in\Z/2}d_{J_f} e^{\sum_f J_f\tilde{\cv}_f\lt[Z(J_f/\l)\rt]+\cdots}\t(J_f/\l)
\ee
where $0\leq\t(j_f)\leq 1$ is a smooth function of compact support located in $j_f\geq o(1)>0$, ``$\cdots$'' stands for   the terms don't scale with $\l$ or of $o(\b^r/\l^s)_{r\leq 2s}$. Since the summand is a compact support function on $\R^{N_f}$, we apply the Poisson resummation formula to the spin-sum:
\be
A(\ck)&=&\lt(\frac{1}{\l}\rt)^{\frac{N_{g,z}}{2}}\sum_{k_f\in\Z}\int_{\R^{N_f}}\lt[d_{J_f} \rmd J_f\rt] e^{\sum_f J_f\lt(\tilde{\cv}_f\lt[Z(J_f/\l)\rt]-4\pi ik_f\rt)+\cdots}\t(J_f/\l)\nonumber\\
&=&\lt(\frac{1}{\l}\rt)^{\frac{N_{g,z}}{2}-2N_f}\sum_{k_f\in\Z}\int_{\R^{N_f}}\lt[j_f\rmd j_f\rt]  e^{\sum_f \l j_f\lt(\tilde{\cv}_f\lt[Z(j_f)\rt]-4\pi ik_f\rt)+\cdots}\t(j_f)\label{universal}
\ee
where $N_f$ denotes the number of triangles in the simplical complex. For each branch $k_f$, we can study the integral use the stationary phase approximation, and we obtain the equation of motion:
\be
\tilde{\cv}_f\lt[Z(j_f)\rt]=4\pi i k_f,\ \  k_f\in\Z
\ee 
where we have used $\frac{\partial\cv_f\lt[j,Z(j)\rt]}{\partial Z}=0$ because of the equations of motion from analytic extended spinfoam action. On the other hand, By Eq.\Ref{real}, $\Re(\cv_f)=0$ implies $\Im(Z(j))=0$, i.e. $Z(j)$ reduces back to the (real) critical data ${g}_{ve},{z}_{vf}$. Then the equation of motion reduces to a restricition of the critical value of $\a_f$
\be
\a_f=4\pi k_f,\ \  k_f\in\Z.
\ee

Such a result can also be derived directly from the original path integral expression of $A(\ck)$:
\be
A(\ck) =\sum_{J_f\in \Z/2}d_{J_f}\int\rmd g_{ve} \int\rmd z_{vf}\ e^{\sum_f J_f\cv_f\lt[g_{ve},z_{vf}\rt]}e^{i\frac{\b}{\l} \sum_{f}J_f \ck_f\lt[g_{ve},z_{vf}\rt]}\t(J_f/\l)
\ee
Again by the Poisson resummation formula:
\be
A(\ck) =(2\l)^{2N_f}\sum_{k_f}\int \lt[j_f\rmd j_f\rt]\int\rmd g_{ve} \int\rmd z_{vf}\ e^{\l \sum_f j_f\lt(\cv_f\lt[g_{ve},z_{vf}\rt]-4\pi i k_f\rt)}e^{i{\b} \sum_{f}j_f \ck_f\lt[g_{ve},z_{vf}\rt]}\t(j_f)
\ee
We employ the stationary phase approximation to analyze the integrals for all branches $k_f$ to obtain the asymptotic expansion in terms of $(\b^r/\l^s)_{r\leq 2s}$ as before. The critical equations $\Re(\cv_f)=0$ and $\delta_g\sum_f j_f\cv_f=\delta_z\sum_f j_f\cv_f=0$ implies the critical configurations classified in Section \ref{criticalconfigurations} are the dominating contributions. At these solutions $\delta_j\sum_f j_f\cv_f=0$ gives the further restriction that the critical values of $\a_f$ have to be
\be
\a_f=4\pi k_f,\ \  k_f\in\Z.
\ee
In the following we assume at the allowed critical values of $j_f$ the compact support function $\t(j_f)=1$.

Although it seems that the expression of spin effective action is not completely necessary in the analysis in this section up to now, the perturbative expression of $W_\ck[j_f,\l,\b]$ is useful later for some certain observations.

\subsection{Effective Degree of Freedom and Effective Amplitude}

Recall Table 1 for the list of critical values of $\a_f$, we find: 
\begin{enumerate}
\item The time-oriented critical configurations of Lorentzian geometry are allowed, and they all contribute the branch $k_f=0$.

\item None of the time-unoriented critical configurations contribute to the leading order since their critical values $\a_f=\eps \pi$ don't satisfy the above critical equation.

\item For the critical configurations of Euclidean and vector geometry, they contribute only when\\ $\eps \lt[\sgn(V_4)\Theta^E_f+\pi n_f\rt]\in 4\pi\Z$ and $\Phi_f=4\pi \Z$.
\end{enumerate}
The critical configurations classified here is the effective degree of freedom in the regime $1\ll\g^{-1}\ll\l\ll\g^{-2}$ from the asymptotic expansion described in the last section.

Therefore we can approximate $A(\ck)$ in the large spin regime $1\ll\g^{-1}\ll\l\ll\g^{-2}$ by summing over all the allowed critical configurations:
\be
A(\ck)=A_{\mathrm{L,O,T}}(\ck)+A_{\mathrm{L,T}}(\ck)+A_{\mathrm{E,V}}(\ck)\label{A1}
\ee
where the expressions of $A_{\mathrm{L,O,T}}(\ck)$, $A_{\mathrm{L,T}}(\ck)$, $A_{\mathrm{E,V}}(\ck)$ are listed in the following (we assume in the following $\eps=1$ fixed by the boundary data, but we suppress the global boundary terms):
\begin{itemize}
\item $A_{\mathrm{L,O,T}}(\ck)$ is given by a sum over all critical configurations $(j_f,g_{ve},z_{ve})_{\mathrm{L,O,T}}$ of globally Lorentzian, Oriented, Time-oriented geometry, with $\sgn(V_4)=\pm1$ globally:
\be
A_{\mathrm{L,O,T}}(\ck)=\lt(\frac{1}{\l}\rt)^{\frac{N_{g,z}-3N_f}{2}}\sum_{(j_f,g_{ve},z_{ve})_{\mathrm{L,O,T}}}j_f e^{i \b\,\sgn(V_4)\sum_f{j}_f{\Theta}_f+o(1)+ o(\b^2/\l)}
\ee

\item $A_{\mathrm{L,T}}(\ck)$ is given by a sum over all critical configurations $(j_f,g_{ve},z_{ve})_{\mathrm{L,T}}$ of globally Lorentzian, Time-oriented geometry. But the geometry is not globally-oriented, i.e. $\sgn(V_4)$ is not a constant:
\be
A_{\mathrm{L,T}}(\ck)=\lt(\frac{1}{\l}\rt)^{\frac{N_{g,z}-3N_f}{2}}\sum_{(j_f,g_{ve},z_{ve})_{\mathrm{L,T}}}{j_f}e^{i \b\sum_{\calr}\sgn(V_4)_\calr\sum_f{j}_f{\Theta}_f+(\mathrm{Boundary Terms})+o(1)+o(\b^2/\l)}
\ee
where $\calr$ denotes the regions in which $\sgn(V_4)$ from $(j_f,g_{ve},z_{ve})_{\mathrm{L,T}}$ is a constant. There is boundary terms in the effective action corresponding to the boundary of each $\calr$, as described in \cite{HZ}.

\item $A_{\mathrm{E,V}}(\ck)$ is given by a sum over the critical configurations $(j_f,g_{ve},z_{ve})_{\mathrm{E,V}}$ of Euclidean and Vector geometry, whose critical values $\a_f\in 4\pi \Z$:
\be
A_{\mathrm{E,V}}(\ck)=\lt(\frac{1}{\l}\rt)^{\frac{N_{g,z}-3N_f}{2}}\sum_{(j_f,g_{ve},z_{ve})_{\mathrm{E,V}}}{j_f}e^{o(1)+o(\b^2/\l)}.
\ee

\end{itemize}

The L.O.T sector $A_{\mathrm{L,O,T}}(\ck)$ of the effective spinfoam amplitude is an analog of quantum Regge calculus, with a discrete functional integration measure, if we ignore $o(\b^2/\l)$ corrections. The functional integration measure is defined on the space of critical configurations $(j_f,g_{ve},z_{vf})$ in the L.O.T sector. By the equivalence theorem in \cite{hanPI}, these critical configurations are equivalent to a set of descrete cotetrad $E_\ell(v)$. Thus $A_{\mathrm{L,O,T}}(\ck)$ can be understood as a functional integration on the space of discrete cotetrad. We suppose $j_f$ is the area of the triangle $f$ measured in the area-unit $a^2_f$, i.e. the area of the triangle is $A_f=j_f a^2_f$. Thus the effective action in $A_{\mathrm{L,O,T}}(\ck)$ can be written as 
\be
i\b\,\sgn(V_4)\sum_f{j}_f\mathring{\Theta}_f=i(\b a^{-2})\,\sgn(V_4)\sum_fA_f{\Theta}_f=\frac{i}{\ell_P^2}\sgn(V_4)\sum_fA_f{\Theta}_f
\ee
where the area-unit relates to the Planck unit by $a^2=\b\ell_P^2=\g\l\ell_P^2$. One may also view that the Regge action and gravitational coupling $\ell_P$ is emergent effectively from the large spin regime of the spinfoam state-sum.

Furthermore, since $\b$ is also a large parameter by $\g^{-1}\ll\l$, we can apply the stationary phase approximation to $A_{\mathrm{L,O,T}}(\ck)$ and obtain an asymptotic expansion in terms of $\b^{-1}$. The Regge action depends on the cotetrad $E_\ell(v)$ only through the edge-lengths $|E_\ell(v)|$. As the leading order contribution in $\b^{-1}$ expansion, the equation of motion satisfied by on-shell $E_\ell(v)$ is nothing but a discrete Einstein equation.

\subsection{Re-expansion of Euclidean and Vector Geometry Sector}\label{reexpansion}

The above results rely on the setting that $\b,\l$ are the only two scaling parameters, which parametrize the perturbation series. Here in this section we show that at the sector of Euclidean and vector geometry critical configurations, we can define another scaling parameter and make an re-expansion of $A_{E.V}(\ck)$, such that some critical configurations with small critical $\a_f$ (mod $4\pi\Z$) can contribute the leading order. Such a strategy is also useful in the analysis of other sectors $A_{L.O.T}(\ck)$ and $A_{L.T}(\ck)$ beyond the regime $\g^{-1}\ll\l\ll\g^{-2}$.   

We develop the perturbation theory from the spinfoam state-sum at a background data $\background$ ($\mathring{J}_f=\l\mathring{j}_f$), which is a critical configuration in $E.V$ sector of the potential $\sum_fj_f\cv_f\lt[g_{ve},z_{vf}\rt]$. We consider the Taylor expansion of the spin effective action $W_\ck[j_f,\l,\b]$ in terms of the spin perturbations $\fs_f=j_f-\mathring{j}_f$:
\be
W_\ck[j_f,\l,\b]&=&i\l\sum_f\mathring{j}_f\mathring{\a}_f+i\l\sum_f\fs_f\mathring{\a}_f+\l\sum_f\mathring{j}_f\sum_Z\frac{\partial\tilde{\cv}_f}{\partial Z}\Big|_{\mathring{g}_{ve},\mathring{z}_{vf}}\frac{\partial Z(j)}{\partial j_f}\Big|_{\mathring{j}_f}\fs_f+\l \sum_{f,f'}\cv_{ff'}\fs_f\fs_{f'}+o(\l\fs^3)\nonumber\\
&+&+i\b\sum_f\mathring{j}_f\sum_Z\frac{\partial\tilde{\ck}_f}{\partial Z}\Big|_{\mathring{g}_{ve},\mathring{z}_{vf}}\frac{\partial Z(j)}{\partial j_f}\Big|_{\mathring{j}_f}\fs_f+i\b \sum_{f,f'}\ck_{ff'}\fs_f\fs_{f'}+o(\b\fs^3)\nonumber\\
&-&\frac{N_{g,z}}{2}\ln\l+\half\ln\det\lt(\frac{2\pi i}{ \sum_fj_f\tilde{\cv}_f''\lt[Z(j)\rt]}\rt)+ o\lt(\frac{\b^r}{\l^s}\rt)_{r\leq 2s}
\ee
where $i\mathring{\a}$ are the critical value of $\cv_f$ at $\background$, and recall that in E.V sector $\mathring{\vth}_f=0$. Since $\background$ is a critical configuration of both the potential $\sum_fj_f\cv_f$ and the spinfoam action $S$, i.e. $\delta_{g}\cv_f=\delta_g S=0$ and $\delta_{z}\cv_f=\delta_z S=0$ at $\background$, therefore we have
\be
\frac{\partial\tilde{\cv}_f}{\partial Z}\Big|_{\mathring{g}_{ve},\mathring{z}_{vf}}=\frac{\partial\tilde{\ck}_f}{\partial Z}\Big|_{\mathring{g}_{ve},\mathring{z}_{vf}}=0.
\ee

We define the following perturbative spin-sum:
\be
&&\sum_{\fs=-\infty}^\infty d_{J_f}\t(\fs)\exp W_\ck[j_f,\l,\b] =\prod_fd_{\mathring{J}_f}e^{i\l\sum_f\mathring{j}_f\mathring{\a}_f}\cz\nonumber\\
&&\cz=\sum_{\fs=-\infty}^\infty \prod_f\lt(1+\frac{2\l\fs_f}{d_{\mathring{J}_f}}\rt)e^{i\l\sum_f\fs_f\mathring{\a}_f+\l \sum_{f,f'}\cv_{ff'}\fs_f\fs_{f'}+i\b \sum_{f,f'}\ck_{ff'}\fs_f\fs_{f'}+\cdots}\t(\fs),
\ee
where $0\leq\t(\fs)\leq 1$ is a smooth function supported on a compact neighborhood of $\mathring{j}_f$. The compact support function $\t(\fs)$ may be viewed as coming from a partition of unity, i.e.
\be
A(\ck)=\sum_{j_f}d_{J_f}A_{J_f}(\ck)=\sum_I\sum_{j_f}d_{J_f}A_{J_f}(\ck)\t_I(j_f)\ \ \ \text{with}\ \ \ \sum_I\t_I(j_f)=1
\ee
where each $\t_I$ are of compact support on a neighborhood $K$ of at most a single critical point (see the following for clarification). If we define $\Fs_f=\l\fs_f=J_f-\mathring{J}_f$ ($\Delta\Fs_f=\half$) as the fluctuation of $J_f$, we can write $\cz$ as
\be
\cz=\sum_{\Fs\in\Z/2} \prod_f\lt(1+\frac{2\Fs_f}{d_{\mathring{J}_f}}\rt)e^{i\sum_f\Fs_f\mathring{\a}_f+\frac{1}{\l} \sum_{f,f'}\cv_{ff'}\Fs_f\Fs_{f'}+i\frac{\b}{\l^2} \sum_{f,f'}\ck_{ff'}\Fs_f\Fs_{f'}+\cdots}\t(\Fs/\l).
\ee
We apply the Poisson resummation formula to $\cz$ and the fact that $2\Fs=2\l\fs\in\Z$
\be
\cz=\sum_{k_f\in\Z}(2\l)^{N_f}\int_{-\infty}^\infty\lt[\rmd\fs_f\rt] \prod_f\lt(1+\frac{2\l\fs_f}{d_{\mathring{J}_f}}\rt)\t(\fs)\, e^{i\l\sum_f\fs_f\lt[\mathring{\a}_f-4\pi k_f\rt]+{\l} \sum_{f,f'}\cv_{ff'}\fs_f\fs_{f'}+i{\b} \sum_{f,f'}\ck_{ff'}\fs_f\fs_{f'}+\cdots}\label{branch}
\ee
where $N_f$ denotes the number of triangles in the simplical complex. 

We denote a short-hand notation:
\be
\a_f(k)\equiv\mathring{\a}_f-4\pi k_f.
\ee
we assume $\a_f(k)$ at the background configuration is small at some certain $k_f$, more precisely we define a new scaling parameter $\eta\ll1$ such that $\a_f(k)\equiv \eta \Fx_f(k)$ with $\Fx_f(k)\sim o(1)$. Then we can define $\xi\equiv \l\eta$ and have
\be
\cz=\sum_{k_f\in\Z}(2\l)^{N_f}\int_{-\infty}^\infty\lt[\rmd\fs_f\rt] \prod_f\lt(1+\frac{2\l\fs_f}{d_{\mathring{J}_f}}\rt)\t(\fs)\, e^{i\xi\sum_f\fs_f\Fx_f(k)+{\l} \sum_{f,f'}\cv_{ff'}\fs_f\fs_{f'}+i{\b} \sum_{f,f'}\ck_{ff'}\fs_f\fs_{f'}+\cdots}
\ee
When we treat $\xi,\l,\b$ are three independent scaling parameters, the stationary phase approximation applied to the integrals obtains the $\l^{-1}$ expansion where
\be
\fs_f=0
\ee
is a solution to the critical equations by $\l$-scaling. The $\l^{-1}$ corrections contains
\be
(\frac{1}{\l})^{s}L_s\lt[\prod_f\lt(1+\frac{2\l\fs_f}{d_{\mathring{J}_f}}\rt)e^{i\xi\sum_f\fs_f\Fx_f(k)}e^{i{\b}\sum_{f,f'}\ck_{ff'}\fs_f\fs_{f'}+\cdots}\t(s)\rt]_{\fs_f=0}
\ee
Since $L_s$ is a differential operator of order $2s$, it gives
\be
\sum_{m+n\leq 2s}\frac{\xi^m\b^n}{\l^s}f_{s,m,n}=\sum_{m+n\leq 2s}{\eta^m\g^n}{\l^{m+n-s}}f_{s,m,n}
\ee
In order that the perturbation expansion is valid, ${\eta^m\g^n}{\l^{m+n-s}}\ll1$ for all terms. A necessary and sufficient condition is 
\be
\eta\ll\l^{-1/2}
\ee

We find that the critical configurations of Euclidean and vector geometry with $\a_f(k)=\eta\Fx_f(k)\ll\l^{-1/2}$ can contribute the leading order in the modified expansion in terms of ${\eta^m\g^n}{\l^{m+n-s}}|_{m+n\leq 2s}$. As a result the re-expansion $A_{E.V}(\ck)$ in Eq.\Ref{A1} is given by a sum over E.V critical configurations satisfying $\a_f(k)\ll\l^{-1/2}$ for some $k_f$:
\be
A_{\mathrm{E,V}}(\ck)=\lt(\frac{1}{\l}\rt)^{\frac{N_{g,z}-3N_f}{2}}\sum_{(j_f,g_{ve},z_{ve})_{\mathrm{E,V}}}{j_f}e^{i\xi\sum_f j_f\Fx_f(k)+\cdots}.
\ee
where $\cdots$ stands for the corrections of $o(1)$ and $o({\eta^m\g^n}{\l^{m+n-s}})_{m+n\leq 2s}$.

Note that the time-unoriented critical configurations of Lorentzian geometry don't contribute even with additional scaling parameter, since the critical value $\a_f=\pi$ doesn't close to any of $4\pi k_f,\ k_f\in\Z$.

\section{Beyond the Regime $\g^{-1}\ll\l\ll\g^{-2}$}

\subsection{$\l^{-1}$-Expansion: Decreasing Degree of Freedom}

The above discussion is restricted in the regime $\g^{-1}\ll\l\ll\g^{-2}$. If $\l\sim o(\g^{-2})$ or even larger, i.e. the spin-sum is located in a regime of spins larger than $\g^{-2}$, in this case the above asymptotic expansion is not valid anymore because $\b^2/\l=\g^2\l$ is not small when $\l\geq \g^{-2}$. Recall the expression of spinfoam action Eq.\Ref{Sj}, if we treat only $\l$ as the scaling parameter, the $\l^{-1}$ expansion can be done for arbitrary value of $\g$, as it is analyzed in \cite{statesum}. In the $\l^{-1}$ expansion, the role played by the critical ${\a}_f$ in Sections \ref{lexp} and \ref{implementation} is now played by the critical value of 
\be
-i\cf_f\equiv -i\cv_f+\g\ck_f,
\ee
see Table 1 for the critical values at different types of critical configurations.

As the previous analysis, we define the spin effective action for the partial amplitude in the large spin regime
\be
A_{J_f}(\ck)=\exp W_\ck[\l j_f]
\ee
By following the same procedure as it it in Section \ref{implementation}, or refer \cite{statesum}, we find that when we make the spin-sum of the partial amplitude $A_{J_f}(\ck)$, the leading contribution in the $\l^{-1}$-expansion is only given by the critical configurations with the critical values of $\cf_f$ satisfying:
\be
\cf_f=\a_f+\g \vth_f\in 4\pi\Z
\ee
Which has been appeared in the literature as the ``flatness problem'' \cite{flatness}. The spinfoam state-sum in the regime $\l\geq o(\g^{-2})$ can be approximated by
\be
A(\ck)=\lt(\frac{1}{\l}\rt)^{\frac{N_{g,z}-3N_f}{2}}\sum_{(j_f,g_{ve},z_{vf})}j_f e^{I(j_f,g_{ve},z_{vf};\l)}\label{A2}
\ee
where $I(j_f,g_{ve},z_{vf};\l)$ contains only $o(\l^{-n})_{n\geq0}$ terms

Here we classify the contributions from different types of critical configurations:
\begin{itemize}

\item For the critical configuration of a globally oriented, time-oriented Lorentz geometry (on the entire triangulation or in a region): 
\be
\text{Bulk:}\ \cf_f=\eps\, \sgn(V_4)\g\Theta_f,\ \ \ \ \text{Boundary:}\ \cf_f=\eps\, \sgn(V_4)\g\Theta^B_f
\ee
where $\Theta_f$ is the deficit angle and $\Theta^B_f$ is the boundary dihedral angle. Therefore the contributions only come from the critical configurations with
\be
\g\Theta_f,\g\Theta_f^B\in 4\pi\Z.
\ee

\item For the critical configuration of a globally oriented, time-unoriented Lorentz geometry (on the entire triangulation or in a region):
\be
\text{Bulk:}\ \cf_f=\eps\lt[ \sgn(V_4)\g\Theta_f+\pi\rt],\ \ \ \text{Boundary:}\ \cf_f=\eps\lt[ \sgn(V_4)\g\Theta^B_f+\pi\rt].
\ee
The contributions come from the critical configurations with
\be
\g\Theta_f,\g\Theta_f^B\in \pm\lt(4\pi \Z-\pi\rt).
\ee

\item For the critical configuration of an Euclidean geometry:
\be
\text{Bulk:}\ \cf_f=\eps\lt[ \sgn(V_4)\Theta^E_f+\pi n_f\rt],\ \ \ \text{Boundary:}\ \cf_f=\eps\lt[ \sgn(V_4)(\Theta^E_f)^{B}+\pi n_f\rt]
\ee
The contributions come from the critical configurations with
\be
\Theta^E_f,(\Theta_f^E)^B\in \pm\lt(4\pi \Z-n_f\pi\rt),\ \ \ n_f=0,1.
\ee

\item For the critical configuration of a vector geometry:
\be
\text{Bulk:}\ \cf_f=\Phi_f,\ \ \ \ \text{Boundary:}\ \cf_f=\Phi^B_f
\ee
where the vector geometry angle $\Phi^B_f$ of the global boundary can be set to be zero by a gauge-fixing of boundary data. The vector geometry contributions come from the critical configurations with
\be
\Phi_f,\Phi^B_f\in 4\pi \Z,\ \ \ n_f=0,1.
\ee

\end{itemize}

Now let's compare the situations in the two different regimes, i.e. $\g^{-1}\ll\l\ll\g^{-2}$ and $\l\geq \g^{-2}$, which are understood respectively in these two different approximation schemes.
\begin{enumerate}
\item A large number of critical configurations of time-oriented Lorentzian geometry, which originally contribute to the leading order in the regime $\g^{-1}\ll\l\ll\g^{-2}$, don't contribute in the regime $\l\geq \g^{-2}$. The survived critical configurations are the ones with $\Theta_f,\Theta_f^B\in 4\pi\Z$.

\item Some critical configurations of time-unoriented Lorentzian geometry, which originally don't contribute to the leading order in $\g^{-1}\ll\l\ll\g^{-2}$, contribute in $\l\geq \g^{-2}$, as $\lt[ \sgn(V_4)\g\Theta_f+\pi\rt]=0$ mod $4\pi\Z$ and $\lt[ \sgn(V_4)\g\Theta^B_f+\pi\rt]=0$ mod $4\pi\Z$.

\item The contribution in the leading order from the critical configurations of Euclidean and vector geometry doesn't change between the $(\b^r/\l^s)_{r\leq 2s}$-expansion in $\g^{-1}\ll\l\ll\g^{-2}$ and the $\l^{-1}$-expansion in $\l\geq \g^{-2}$, essentially because the critical value of $\ck_f$ vanishes for both types of critical configurations.
\end{enumerate}

From the above classification, we find there is a significant drop-off of the effective degrees of freedom in the leading order approximation of $A(\ck)$ (from Eq.\Ref{A1} to Eq.\Ref{A2}), when the large spin regime changes from $\g^{-1}\ll\l\ll\g^{-2}$ to $\l\geq \g^{-2}$. The significant decreasing of the degrees of freedom mainly comes from the critical degrees of freedom corresponding to the time-oriented Lorentzian geometry. Such a dramatic change of degrees of freedom may hint to a possible phase transition.

There exists another point of view: Essentially the small Barbero-Immirzi parameter $\g\ll1 $ implies the existence of the two large spin regimes $\g^{-1}\ll\l\ll\g^{-2}$ and $\l\geq \g^{-2}$ with different effective degrees of freedom. If $\g$ is tuned larger so that $\g\sim o(1)$, the regime $\g^{-1}\ll\l\ll\g^{-2}$ is not a large spin regime anymore. Then the large spin effective degree of freedom as $\g\sim o(1)$ is only the same as the ones originally in the regime $\l\geq \g^{-2}$. Therefore the number of effective degrees of freedom in the large spin regime depends on the value of the Barbero-Immirzi parameter.

\subsection{Re-expansion: Detailed Decreasing Behavior}\label{decreasing}

Recall the strategy in Section \ref{reexpansion}, when we consider a re-expansion of $A_{E.V}(\ck)$ by introducing a new scaling parameter, such a strategy can be applied to other sectors in the regime $\l\geq \g^{-2}$, in order to obtain some knowledge about the detailed decreasing behavior of the effective degree of freedom.

The perturbative expression of $W_{\ck}[\l j_f]$ can be obtained by employing Theorem \ref{almostanalytic} (see \cite{statesum} for details):
\be
W_{\ck}[\l j_f]=i\l\sum_f \mathring{j}_f\mathring{\cf}_f+i\l\sum_f\mathring{\cf}_f\fs_f+\l\sum_{f,f'}W_{f,f'}\fs_f\fs_{f'}+o(\l\fs^3)-\frac{N_{g,z}}{2}\ln\l+\cdots
\ee
where $\mathring{\cf}_f$ is the critical value of $\cf_f$ at the background data $\background$. Here ``$\cdots$'' stands for the corrections of $o(1)$ and $o(\l^{-1})$. 

As it is done in Section \ref{reexpansion}, we define the following perturbative spin-sum and make the Poisson resummation:
\be
&&\sum_{\fs=-\infty}^\infty d_{J_f}\t(\fs)\exp W_\ck[\l j_f] =\prod_fd_{\mathring{J}_f}e^{i\l\sum_f\mathring{j}_f\mathring{\cf}_f}\cz\nonumber\\
&&\cz=\sum_{k_f\in\Z}(2\l)^{N_f}\int_{-\infty}^\infty\lt[\rmd\fs_f\rt] \prod_f\lt(1+\frac{2\l\fs_f}{d_{\mathring{J}_f}}\rt)\t(\fs)\, e^{i\l\sum_f\fs_f\lt[\mathring{\cf}_f-4\pi k_f\rt]+{\l} \sum_{f,f'}W_{ff'}\fs_f\fs_{f'}+\cdots}
\ee
where $N_f$ denotes the number of triangles in the simplical complex, $\t(\fs)$ is a smooth function of compact support coming from a partition of unity. 

We define a new scaling parameter by 
\be
\mathring{\cf}_f-4\pi k_f\equiv \eta\Fx_f(k)\ \ \ \text{and}\ \ \ \xi\equiv \l\eta 
\ee
for some certain $k_f$, and assuming $\eta\ll 1$. Then $\cz$ can be written as
\be
\cz=\sum_{k_f\in\Z}(2\l)^{N_f}\int_{-\infty}^\infty\lt[\rmd\fs_f\rt] \prod_f\lt(1+\frac{2\l\fs_f}{d_{\mathring{J}_f}}\rt)\t(\fs)\, e^{i\xi\sum_f\fs_f\Fx_f(k)+{\l} \sum_{f,f'}W_{ff'}\fs_f\fs_{f'}+\cdots}
\ee
Applying the stationary phase approximation by the $\l$-scaling, we have a $\l^{-1}$-expansion with the leading contribution given by $\fs_f=0$. But we know that such an expansion is actually in terms of $(\xi^r/\l^s)_{r\leq 2s}$, which is valid only when $\xi\ll \l^{1/2}$ i.e. $\eta\ll \l^{-1/2}$.   

The resulting approximation of spinfoam state-sum is given by
\be
A(\ck)=\lt(\frac{1}{\l}\rt)^{\frac{N_{g,z}-3N_f}{2}}\sum_{(j_f,g_{ve},z_{ve})}{j_f}e^{i\xi\sum_f j_f\Fx_f(k)+\cdots}.
\ee
where $\cdots$ stands for the corrections of $o(1)$ and $o(\eta^r\l^{r-s})_{r\leq 2s}$. The above sum is over all the critical configurations satisfying ${\cf}_f-4\pi k_f\equiv \eta\Fx_f(k)\ll o(\l^{-1/2})$ for some $k_f$.

Let's consider as an example the contribution from the globally oriented and time-oriented critcal configurations of Lorentzian geometry, where the most significant decreasing of degree of freedom happens. At these critical configurations $\cf_f=\g\Theta_f$, in the regime $\l\geq \g^{-2}$ the effective degree of freedom is given by the critical configurations satisfying
\be
\g\Theta_f=\eta\Fx_f(k)\ll o(\l^{-1/2})
\ee
in the $k_f=0$ branch as an example. It results in the the following bound for the allowed deficit angle:
\be
|\Theta_f|\ll \g^{-1}\l^{-1/2}
\ee 
As $\l\geq \g^{-2}$, the deficit angle has to satisfy $|\Theta_f|\ll1$. As $\l$ increases, the above bound becomes smaller so that $\Theta_f\to 0$ asymptotically as $\l\to\infty$.

Note that in this case since the new scaling parameter $\eta$ essentially parametrizes the small deficit angle, the expansion in terms of $(\eta^r\l^{r-s})_{r\leq 2s}$ may be viewed as a curvature expansion.

\section{Discussion}

The analysis in the present paper take into account of the sum over spins in the semiclassical/asymptotic analysis of spinfoam state-sum model. We show that in the large spin regime of the spinfoam state-sum can be divide into two separate regimes $\g^{-1}\ll\l\ll\g^{-2}$ and $\l\geq \g^{-2}$, provided that the Barbero-Immirzi parameter is small $\g\ll 1$. The spinfoam state-sum amplitude behaves differently in these two regimes. In the regime $\g^{-1}\ll\l\ll\g^{-2}$, the critical configurations $(j_f,g_{ve},z_{vf})$ of Lorentzian geometry with arbitrary deficit/dihedral angles $\Theta_f$ can contribute in the leading order, while most of them doesn't contribute to the leading order in the regime $\l\geq \g^{-2}$, unless their deficit/dihedral angles satisfies $|\g\Theta_f|\ll \l^{-1/2}$ mod $4\pi\Z$. 

Interestingly in the regime $\g^{-1}\ll\l\ll\g^{-2}$, the leading contribution of the spinfoam state-sum is given in Eq.\Ref{A1}. Only the time-oriented critical configurations of Lorentzian geometry contribute in the leading order among all the Lorentzian critical configurations. The leading contribution given by them is an analog of quantum Regge calculus with a discrete functional integration measure.

Moreover, when the background spin parameter $\l$ increases from $\g^{-1}\ll\l\ll\g^{-2}$ and $\l\geq \g^{-2}$, we observe that there is a significant decrease of effective degrees of freedom, mainly coming from the sector of critical configurations of time-oriented Lorentzian geometry. Such a result seems to hint the possibility that there may exists a phase transition of certain type between the two regimes.

An order parameter is needed in order to parametrize the possible phase transition between $\g^{-1}\ll\l\ll\g^{-2}$ and $\l\geq \g^{-2}$. As a candidate of order parameter, we choose the expectation value of $\cf_f\lt[g_{ve},z_{vf}\rt]$ where the spinfoam action is written as $S=\sum_fJ_f\cf_f$. The expectation value is given by
\be
\lag\cf_f\rag=\frac{\sum_{j_f}d_{J_f}\int\rmd g_{ve}\rmd{z}_{vf}\,e^{\l S\lt[j_f,g_{ve},z_{vf}\rt]}\cf_f\lt[g_{ve},z_{vf}\rt]}{\sum_{j_f}d_{J_f}\int\rmd g_{ve}\rmd{z}_{vf}\,e^{\l S\lt[j_f,g_{ve},z_{vf}\rt]}}
\ee
As a candidate of order parameter, we need to compare the behavior of $\lag\cf_f\rag$ in two different large spin regimes $\g^{-1}\ll\l\ll\g^{-2}$ and $\l\geq \g^{-2}$. To simplify the problem, we only considering the sector of critical configurations with their critical $\cf_f$ relatively small and not close to $4\pi$, i.e. the $k_f=0$ branch.

\begin{itemize}

\item In the regime $1\ll\g^{-1}\ll\l\ll\g^{-2}$, by the approximation toward Eq.\Ref{A1}, we obtain
\be
\lag\cf_f\rag=\lag\cf_f\rag_{\mathrm{L.O.T}}+\lag\cf_f\rag_{\mathrm{L.T}}+\lag\cf_f\rag_{\mathrm{E.V}}
\ee
Both $\lag\cf_f\rag_{\mathrm{L.O.T}}$ and $\lag\cf_f\rag_{\mathrm{L.T}}$ are of $o(1)$ since they are analogs of averaging deficit angle in quantum Regge calculus, while $\lag\cf_f\rag_{\mathrm{E.V}}\ll o(\l^{-\half})$. Therefore $\lag\cf_f\rag\sim o(1)$ in the regime $1\ll\g^{-1}\ll\l\ll\g^{-2}$.

\item However $\lag\cf_f\rag\ll o(\l^{-\half})$ in the regime $\l\geq \g^{-2}$ by the analysis in Section \ref{decreasing}.

\end{itemize}

\begin{figure}[h]
\begin{center}
\includegraphics[width=10cm]{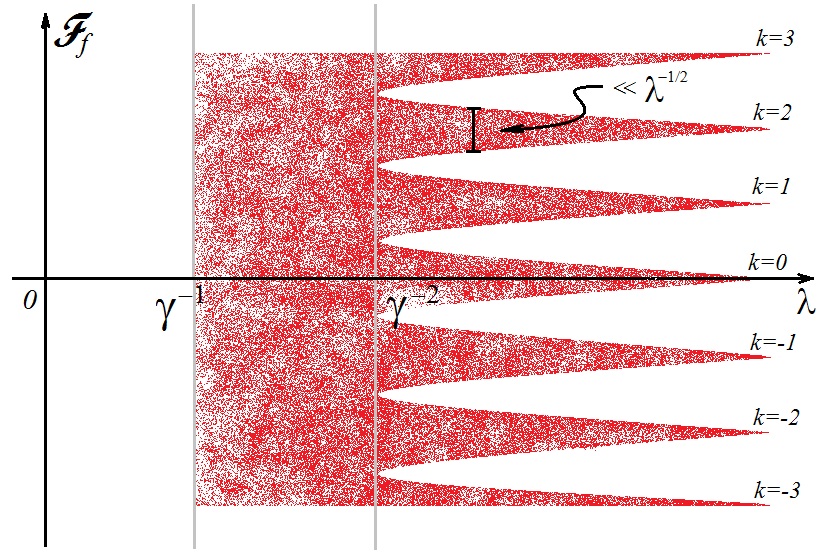}
\caption{In two different regimes $\g^{-1}\ll\l\ll\g^{-2}$ and $\l\geq \g^{-2}$, the figure illustrates the allowed spinfoam critical configurations, which contribute the leading order approximation of $A(\ck)$. the critical configurations are (partially) parametrized by $\l$ and the critical values of $\cf_f$ of an arbitrary triangle $f$. In the regime $\g^{-1}\ll\l\ll\g^{-2}$, the critical configurations with arbitrary $\cf_f$ are allowed, while only the ones with $|\cf_f-4\pi k_f|\ll \l^{-1/2}$ are allowed in the regime $\l\geq \g^{-2}$.}
  \label{phases}
  \end{center}
\end{figure}

\begin{figure}[h]
\begin{center}
\includegraphics[width=10cm]{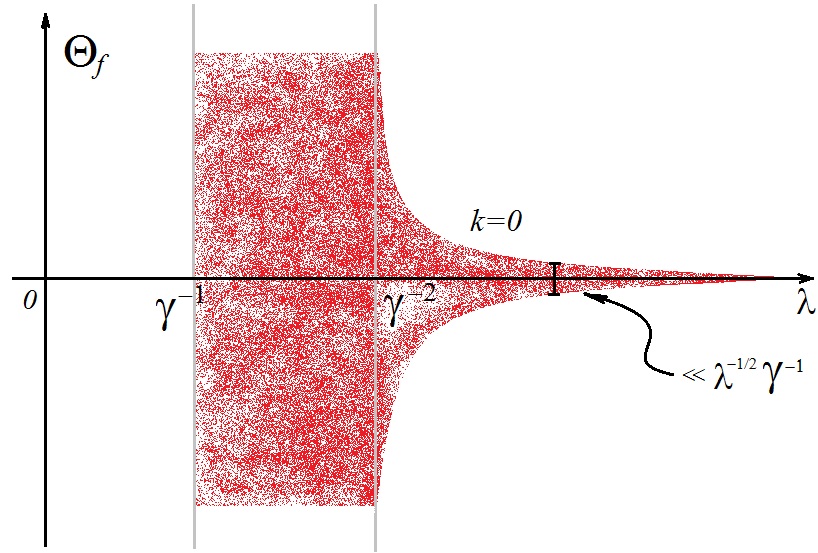}
\caption{The figure shows the situation corresponding to $k_f=0$ branch only. In two different regimes $\g^{-1}\ll\l\ll\g^{-2}$ and $\l\geq \g^{-2}$, the figure illustrates the allowed time-oriented critical configurations of Lorentzian geometry, which contribute the leading order approximation of $A(\ck)$. the critical configurations are (partially) parametrized by $\l$ and the deficit angle $\Theta_f$ of an arbitrary triangle $f$. In the regime $\g^{-1}\ll\l\ll\g^{-2}$, the time-oriented Lorentzian critical configurations with arbitrary $\Theta_f$ are allowed, and they all contribute the $k_f=0$ branch, while only the ones with $|\Theta_f|\ll \l^{-1/2}\g^{-1}$ (in $k_f=0$ branch) are allowed in the regime $\l\geq \g^{-2}$.}
  \label{phases1}
  \end{center}
\end{figure}

Finally we illustrate the situation of the effective degrees of freedom in the two regimes by FIGs.\ref{phases} and \ref{phases1}. Both figures draw the space of spinfoam configurations $(J_f,g_{ve},z_{vf})$ when $J_f=\l j_f$ with $\l\gg1$. The red points illustrate the allowed spinfoam critical configurations, which contribute the leading order approximation of $A(\ck)$. In FIG.\ref{phases} all types of the critical configurations are shown with all branches $k_f$, while FIG.\ref{phases1} only shows the time-oriented critical configurations of Lorentzian geometry with $k_f=0$ branch only.

\section*{Acknowledgements}

The author would like to thank H. Haggard, S. Speziale, A. Riello, C. Rovelli, and M. Zhang for many helpful discussions. He also would like to thank Y. Ma for the invitation to visit the Center for Relativity and Gravitation, Beijing Normal University, where a part of this research work is carried out. The research leading to these results has received funding from the People Programme (Marie Curie Actions) of the European Union's Seventh Framework Programme (FP7/2007-2013) under REA grant agreement No. 298786.


\begin{thebibliography}{20}



\bibitem{book}T. Thiemann. \emph{Modern Canonical Quantum General Relativity} (Cambridge University Press, Cambridge, 2007)\\
C. Rovelli. \emph{Quantum Gravity} (Cambridge University Press 2004)

\bibitem{rev}A. Ashtekar and J. Lewandowski. Background independent quantum gravity: A status
report. {Class. Quant. Grav.} {21} (2004) R53.\\
M. Han, W. Huang and Y. Ma. Fundamental structure of loop quantum gravity. Int. J. Mod. Phys.
D16 (2007) 1397-1474 [arXiv:gr-qc/0509064].

\bibitem{sfrevs}C. Rovelli. Zakopane lectures on loop gravity. [arXiv:1102.3660]\\
A. Perez. The spin foam approach to quantum gravity. Living Rev. Relativity 16 (2013) 3

\bibitem{EPRL}
J. Engle, E. Livine, R. Pereira and C. Rovelli. LQG vertex with finite Immirzi parameter. {Nucl.
Phys.} B{799} (2008) 136

\bibitem{Carlo}C. Rovelli. Simple model for quantum general relativity from loop quantum gravity. J. Phys. Conf. Ser. 314 (2011) 012006

\bibitem{semiclassical}J. W. Barrett, R. J. Dowdall, W. J. Fairbairn, F. Hellmann and R. Pereira. Lorentzian spin foam amplitudes: graphical calculus and asymptotics. Class. Quant. Grav. 27 (2010) 165009

\bibitem{CF}F. Conrady and L. Freidel. On the semiclassical limit of 4d spin foam models. Phys. Rev. D78 (2008) 104023

\bibitem{HZ}M. Han and M. Zhang. Asymptotics of spinfoam amplitude on simplicial manifold: Euclidean theory. Class. Quantum Grav. 29 (2012) 165004 [arXiv:1109.0500]\\
M. Han and M. Zhang. Asymptotics of spinfoam amplitude on simplicial manifold: Lorentzian theory. [arXiv:1109.0499]

\bibitem{hanPI}M. Han and T. Krajewski. Path Integral Representation of Lorentzian Spinfoam Model, Asymptotics, and Simplicial Geometries. [arXiv:1304.5626]

\bibitem{flatness}F. Hellmann and W. Kaminski. Geometric asymptotics for spin foam lattice gauge gravity on arbitrary triangulations. [arXiv:1210.5276]\\
C. Perini. Holonomy-flux spinfoam amplitude. [arXiv:1211.4807]\\
V. Bonzom. Spin foam models for quantum gravity from lattice path integrals. Phys. Rev. D80 (2009) 064028

\bibitem{claudio}E. Magliaro and C. Perini. Emergence of gravity from spinfoams. Europhys. Lett. 95 (2011) 30007\\
E. Magliaro and C. Perini. Regge gravity from spinfoams. [arXiv:1105.0216]

\bibitem{QSF}M. Han. 4-dimensional spinfoam model with quantum Lorentz group. J. Math. Phys. 52 (2011) 072501 [arXiv:1012.4216]\\
W. J. Fairbairn and C. Meusburger. Quantum deformation of two four-dimensional spin foam models. J. Math. Phys. 53 (2012) 022501

\bibitem{QSFasymptotics}M. Han. Cosmological constant in LQG vertex amplitude. Phys. Rev. D 84 (2-11) 064010 [arXiv:1105.2212]\\
S. Mizoguchi and T. Tada.  3-dimensional gravity from the Turaev-Viro invariant. Phys. Rev. Lett. 68 (1992) 1795-1796

\bibitem{statesum}M. Han. On spinfoam model in large spin regime. [arXiv:1304.5627]


\bibitem{propagator}E. Bianchi and Y. Ding. Lorentzian spinfoam propagator. Phys. Rev. D 86 (2012) 104040\\
E. Bianchi, E. Magliaro, and C. Perini. LQG propagator from the new spin foams. Nucl. Phys. B822 (2009) 245-269

\bibitem{3pt}C. Rovelli and M. Zhang. Euclidean three-point function in loop and perturbative gravity. Class. Quantum Grav. 28 (2011) 175010 [arXiv:1105.0566]

\bibitem{stationaryphase}L. H\"ormander. \emph{The analysis of linear partial differential operators I. Distribution theory and Fourier analysis}, second edition. 1990, Springer-Velag, Berlin

\bibitem{almostanalytic}A. Melin and J. Sj\"ostrand. Fourier integral operators with complex-valued phase functions. Lecture Notes in Mathematics 459 (1975) 120-223 






















\end{thebibliography}
\end{document}